\documentclass[prb,onecolumn,preprintnumbers,amsmath,amssymb,floatfix,showpacs]{revtex4}

\usepackage[dvips]{graphics,graphicx,color}  \newcommand{\vect}[1]{{\bf {#1}}}

\begin{document}

\title{Diamond$\rightarrow \! \beta$-tin phase transition in Si within
diffusion quantum Monte Carlo}

\author{Ryo Maezono}

\affiliation{Japan Advanced Institute of Science and Technology, School of
Information Science, Asahidai 1-1, Nomi, Ishikawa 923-1292, Japan}

\author{N.\ D.\ Drummond, A.\ Ma, and R.\ J.\ Needs}

\affiliation{TCM Group, Cavendish Laboratory, University of Cambridge, J.\ J.\
Thomson Avenue, Cambridge CB3 0HE, United Kingdom}

\begin{abstract} We have studied the diamond$\rightarrow \! \beta$-tin
  phase transition in Si using diffusion quantum Monte Carlo (DMC) methods.
  Slater-Jastrow-backflow trial wave functions give lower DMC energies than
  Slater-Jastrow ones, and backflow slightly favors the $\beta$-tin phase with
  respect to the diamond phase.  We have investigated the changes in the
  equation of state that result from the use of different pseudopotentials,
  the inclusion of either zero-point motion or finite-temperature vibrations,
  and the application of corrections for finite-size effects.  Our tests
  indicate that the choice of pseudopotential can significantly affect the
  equation of state.  Using a Dirac-Fock pseudopotential leads to an
  overestimation of the transition pressure but an empirical pseudopotential
  designed for use in correlated calculations gives a transition pressure in
  quite good agreement with experiment.
\end{abstract}





\maketitle

\section{Introduction \label{sec:introduction}}

Calculating accurate energy differences between the atomic structures of
materials is one of the central problems in quantum mechanical studies of
condensed matter.  Here we study the first-order pressure-induced structural
phase transition between the diamond and $\beta$-tin phases of Si.  This is a
demanding problem because of the substantial volume change and the increase in
the coordination number from four to six at the transition, and the fact that
the structural phase transition is accompanied by a metal-insulator
transition.  Many theoretical studies of the diamond$\rightarrow \! \beta$-tin
phase transition in Si have been reported, mainly using first-principles
density-functional theory (DFT)
methods.\cite{Yin_1982,Needs_1984,boyer91,needs95,moll_1995,Gaal-Nagy_2001,mujica03,Gaal-Nagy_2004}
The DFT results show significant dependence on the density functional
used,\cite{batista_2006} indicating the necessity for an accurate description
of exchange-correlation effects.  A similar dependence on the functional has
been noted for self-interstitial defects in
Si.\cite{leung99,batista_2006,Mattsson_2008} These results motivate studies
using other descriptions of many-body effects such as that provided by quantum
Monte Carlo (QMC) methods.

Pressure-induced phase transitions have been studied using the diffusion
quantum Monte Carlo (DMC)\cite{ceperley_1980,foulkes_2001} and auxiliary-field
quantum Monte Carlo (AFQMC) methods.\cite{Zhang_2003} The pressure-induced
phase transition of Si from the diamond to the $\beta$-tin structure has been
studied previously using both DMC\cite{alfe_2004_betatin,hennig_2010} and
AFQMC,\cite{Purwanto_2009} while DMC studies of structural phase transitions
have also been reported for N$_2$,\cite{Mitas_nitrogen} MgO,\cite{Alfe_2005}
FeO,\cite{Kolorenc_2008} SiO$_2$,\cite{silica_qmc_2010} and
BN\@.\cite{bn_qmc_2010}

The purpose of the present study is to test recent developments in DMC
methodology.  We have used Slater-Jastrow-backflow (SJB) trial wave functions
which go beyond the single-particle nodal surfaces of Slater-Jastrow (SJ) wave
functions and significantly lower the energies of the diamond and $\beta$-tin
phases.  We have tested two recent and very different schemes for calculating
corrections to the DMC energies obtained with finite supercells that allow us
to extrapolate to the infinite-cell limit.  To compare the DMC transition
pressure and volumes with experiment it is necessary to include the
vibrational free energy, which we have re-calculated within DFT with greater
accuracy than was achieved in previous studies.  Finally we have calculated
the energy-volume relations within DFT using different pseudopotentials in
order to study the likely influence of the pseudopotential on the DMC
transition pressure and volumes and the equations of state of the diamond and
$\beta$-tin structures.

The rest of the paper is organized as follows.  In Sec.\
\ref{subsec:hamiltonian} we define the Hamiltonians used and in Sec.\
\ref{subsec:trial_wave_functions} the trial wave functions are described.
Finite-size corrections are discussed in Sec.\
\ref{subsec:finite_size_effects} and the phonon calculations are described in
Sec.\ \ref{subsec:phonon calculations}.  The main DMC results are reported in
Sec.\ \ref{subsec:DMC results}.  Section \ref{subsec:testing pseudopotentials}
describes our method for testing different pseudopotentials and reports and
discusses results for the equation-of-state parameters and transition
pressures and volumes.  We draw our conclusions in Sec.\
\ref{sec:conclusions}. The QMC calculations were performed with the
\textsc{casino} code,\cite{casino_ref} and for the DFT calculations we used
the \textsc{castep} plane-wave pseudopotential code.\cite{ClarkSPHPRP05}

\section{The Hamiltonian \label{subsec:hamiltonian}}

We used the Dirac-Fock pseudopotential of Trail and Needs\cite{Trail_2005} to
describe the interaction of the valence electrons with the Si$^{4+}$ ions.  We
also added a core-polarization potential (CPP) to each
ion.\cite{shirley93,lee03} In the CPP approximation, the polarization of each
core is determined by the electric field at the nucleus from the instantaneous
positions of the electrons and the other atomic cores.  CPPs therefore give an
approximate description of both dynamical core-valence correlation and static
polarization effects.  We used the CPP parameters from Ref.\
\onlinecite{shirley93} and the CPP energy was evaluated using the scheme
described in Ref.\ \onlinecite{maezono_2003}. The non-local pseudopotential
energy was calculated using the variational scheme of Ref.\
\onlinecite{Casula_2006}.

We used 54-atom simulation cells subject to periodic boundary conditions.  To
make the simulation cell for the diamond structure we took the standard set of
primitive translation vectors of the two-atom face-centered-cubic cell and
multiplied them by three to give a cell accommodating 54 atoms.  The
$\beta$-tin structure has a body-centered tetragonal lattice and a two-atom
primitive unit cell.  We used a $c/a$ ratio of 0.556 for the QMC calculations,
which is close to the experimental value of 0.5498 reported at the phase
transition.\cite{McMahon_1994} The energy difference between the structures
when using these two $c/a$ ratios is negligible on the scale of interest.  The
primitive lattice vectors of the two-atom $\beta$-tin structure can be chosen
to be ${\bf u} = (a,a,c)$, ${\bf v} = (a,-a,c)$, and ${\bf w} = (a,a,-c)$.  We
wrote a code to search over all possible 54-atom cells constructed from linear
combinations of ${\bf u}$, ${\bf v}$, and ${\bf w}$ and chose the cell with
the maximum value of the shortest distance between supercell lattice points,
which has translation vectors ${\bf U}$, ${\bf V}$, and ${\bf W}$, where ${\bf
U} = -9{\bf u} + 3{\bf v} + 6{\bf w}$, ${\bf V} = 3{\bf v}$, and ${\bf W} =
-3{\bf u} +2{\bf v} + {\bf w}$. Using such a simulation cell gives the largest
separation between each electron and its periodic image, which helps to reduce
the finite-size effects arising from long-ranged exchange-correlation effects
and allows the isotropic terms in the Jastrow factor and backflow
transformation (see Sec.\ \ref{subsec:trial_wave_functions}) to give a good
description of the electron correlation.  The Ewald potential was used to
model the interactions between point charges.

\section{Trial wave functions \label{subsec:trial_wave_functions}}

The DMC method relies on the availability of an accurate trial wave function
which is used to guide the sampling and to enforce the fixed-node
constraint.\cite{ceperley_1980,foulkes_2001} We used trial wave functions of
SJ and SJB types.  The Slater-Jastrow form is
\begin{equation}
\label{eq:slater-jastrow} \Psi_{\rm SJ}({\bf R})\! \! = \exp[J({\bf R})]
\det{\left[ \psi_n({\bf r}_i^{\uparrow})\right]} \det{\left[ \psi_n({\bf
r}_j^{\downarrow})\right]} \;,
\end{equation} where ${\bf R}$ denotes the positions of all the
electrons, ${\bf r}_i^{\uparrow}$ is the position of the $i$th spin-up
electron, ${\bf r}_j^{\downarrow}$ is the position of the $j$th spin-down
electron, $\exp[J({\bf R})]$ is the Jastrow factor, and $\det{\left[
\psi_n({\bf r}_i^{\uparrow}) \right]}$ and $\det{\left[ \psi_n({\bf
r}_i^{\downarrow}) \right]}$ are determinants of single-particle orbitals.

The orbitals were calculated using the \textsc{castep} code and a large
basis-set energy cutoff of 1088 eV\@.  The orbitals were then transformed into
a B-spline or ``blip'' polynomial basis for greater
efficiency.\cite{alfe_2004_blips} We used Jastrow factors consisting of
polynomial electron-nucleus (en) and electron-electron (ee) terms and a
plane-wave ee term,\cite{ndd_jastrow} with a total of 26 optimizable
parameters.  We used wave functions both with and without backflow
transformations.\cite{backflow} In the SJB wave function the particle
coordinates $\vect{\bf r}_i$ in the determinants in Eq.\
(\ref{eq:slater-jastrow}) are replaced by collective coordinates
\begin{equation} \vect{x}_i(\vect{R})=\vect{r}_i+\mbox{\boldmath
$\xi$}_i(\vect{R}) \;,
\end{equation} where $\mbox{\boldmath $\xi$}_i(\vect{R})$ is the
backflow displacement of electron $i$, which depends on the positions of all
of the electrons and ions.  The backflow displacement consisted of polynomial
ee and en terms\cite{backflow} and contained a total of 24 optimizable
parameters.  The wave-function parameters were optimized by a variational
Monte Carlo (VMC) procedure in which we first minimized the variance of the
energy\cite{umrigar_1988a,ndd_newopt} and then minimized the variational
energy itself.\cite{umrigar_emin} The wave function parameters were optimized
at a single ${\bf k}_s$ point in the Brillouin zone of the simulation cell
(this being $L$ and $M$ for the diamond and $\beta$-tin structures,
respectively) and the same Jastrow and backflow parameters were subsequently
used for each ${\bf k}_s$ point studied, where ${\bf k}_s$ is the many-body
Bloch vector.\cite{rajagopal_kpoints}

All of the results reported here were obtained using a DMC time step $\Delta
\tau$ of 0.01 a.u.  Some test calculations using a SJ trial wave function were
also performed with $\Delta \tau = $0.02 a.u., which give very similar results
to those with $\Delta \tau = $0.01 a.u.  We used a target population of 640
configurations in each of our DMC calculations.

\section{Finite-size corrections \label{subsec:finite_size_effects}}

The 54-atom simulations give a reasonable description of the infinite system,
but it is still necessary to make finite-size corrections to obtain accurate
results.  The finite-size effects can be divided into a part which is very
similar to the Brillouin zone integrations performed in single-particle
calculations, and exchange-correlation effects arising from electron
separations beyond the simulation cell.  Brillouin zone integration is more
important in metals, where it is required to describe Fermi-surface effects,
than in insulators, where a sparse sampling of the Brillouin zone is
sufficient.  Our 54-atom cells are equivalent to sampling the Brillouin zone
of the two-atom primitive cell in a single-particle calculation with 27 ${\bf
k}$ points.

It is only possible to perform calculations with one simulation-cell Bloch
vector ${\bf k}_s$ at a time within explicitly correlated methods.  For the
insulating diamond phase we used only the $L$-point, which allows the use of a
real trial wave function while giving a good description of the energy and
charge density in single-particle calculations.\cite{rajagopal_kpoints,kent99}
For the metallic $\beta$-tin phase we performed DMC calculations at twelve
randomly chosen ${\bf k}_s$ points and averaged the results.\cite{lin_twist_av}

We tested two different schemes for calculating additional finite-size
corrections.  In the first scheme we added a further ${\bf k}$-point
correction consisting of the difference between the PBE-DFT energies with a
very dense ${\bf k}$-point sampling and with the finite ${\bf k}$-point
sampling corresponding to that used in the QMC calculations.  We then added a
correction arising from the long-ranged exchange-correlation interaction using
the method described in Refs.\ \onlinecite{fin_chiesa} and
\onlinecite{Drummond_2008}, which involves evaluating the static structure
factor and two-body Jastrow factor on the grid of reciprocal lattice vectors
within QMC and interpolating to obtain an approximate static structure factor
and two-body Jastrow factor for the infinite system, from which a finite-size
correction can be evaluated.  The second scheme is that of Kwee \textit{et
al.}\cite{Kwee_2008} which involves performing a DFT calculation using a
modified local density approximation (LDA) functional obtained from DMC
calculations for homogeneous electron gases in finite simulation cells subject
to periodic boundary conditions.  This method includes both the ${\bf
k}$-point and long-ranged exchange-correlation corrections, but it also
involves making the somewhat stronger assumptions that the LDA is sufficiently
accurate and the correction is insensitive to the shape of the simulation cell.

We found excellent agreement between the finite-size corrections obtained from
the structure factor method\cite{fin_chiesa,Drummond_2008} and the DFT method
of Kwee \textit{et al.}\cite{Kwee_2008} The finite-size corrections themselves
are quite large, amounting to an increase in energy of about 0.11 eV per atom
for the diamond structure and about 0.099 eV per atom for $\beta$-tin.
However, the difference between the finite-size corrections obtained from the
two schemes is, in the worst case, about 0.007 eV per atom, while the average
of the absolute deviation over all of the data points for the two structures
is 0.004 eV per atom.

\begin{figure}
\includegraphics[clip,scale=0.3]{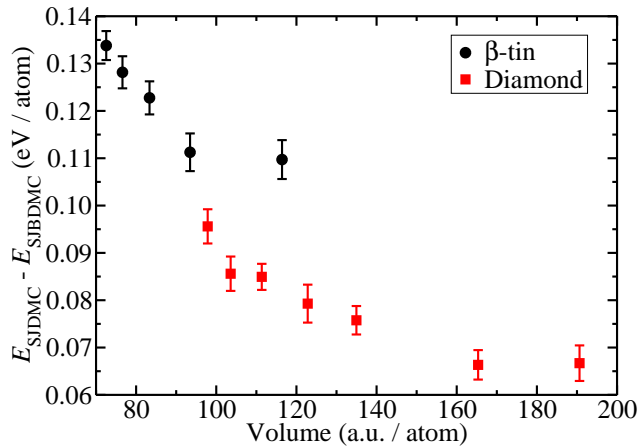}
\caption{(Color online) Difference between the SJ and SJB DMC energies for the
diamond and $\beta$-tin phases.  The diamond- and $\beta$-tin-structure
calculations were performed at the L and M points in the Brillouin zone,
respectively. \label{fig:backflow}}
\end{figure}

\section{Phonon calculations \label{subsec:phonon calculations}}

We evaluated the phonon contribution to the Helmholtz free energy of each
phase as a function of volume using the PBE functional.  We used a 64-atom
supercell, an energy cutoff of 653 eV, and a $6\times6\times6$ ${\bf k}$-point
mesh to calculate the phonons of the $\beta$-tin structure and a 54-atom
supercell, a 435 eV cutoff, and a $3 \times 3 \times 3$ ${\bf k}$-point grid
for the diamond structure.  The phonons frequencies and modes were calculated
using a finite-displacement method and the results were extrapolated to zero
displacement.  Ga\'al-Nagy \textit{et al.}\cite{gaal-nagy_2008} have shown
that it is difficult to obtain stable phonon modes of the $\beta$-tin phase of
Si in DFT calculations.  Our study supports this conclusion, and we have found
stable phonon modes only by carefully relaxing the $c/a$ ratio at every volume
and, even then, the phonon modes are only stable over the volume range
91.4--111.8 a.u.\ per atom.  Over the volume range where the phonons are
stable, the variation of the relaxed $c/a$ ratio with volume is well fitted by
$c/a = 0.521075+0.377765/(131.41-V)^{0.77666}$, where $V$ is the volume per
atom in a.u.

The phonon free energy was calculated from the phonon frequencies.  The
previous DMC and AFQMC calculations estimated the changes in the transition
pressure from nuclear zero-point motion and finite-temperature vibrations
using the DFT vibrational data of Ga\'al-Nagy \textit{et
al.}\cite{gaal-nagy99} These data suggest that including zero-point motion
reduces the phase transition pressure by about 1 GPa, and that
finite-temperature effects at 300 K reduce the transition pressure by a
further 0.3 GPa.  We believe that Ga\'al-Nagy \textit{et
al.}\cite{gaal-nagy99} have somewhat overestimated the size of the vibrational
effects because they used a non-optimal $c/a$ ratio.  Our phonon calculations
give a reduction in the transition pressure from zero point motion of about
0.62 GPa and a further reduction of 0.34 GPa at 300 K.  These values do not
differ significantly when we use LDA, PBE, or DMC static-lattice energy-volume
curves.

\section{DMC results \label{subsec:DMC results}}

The reduction in the DMC energy resulting from the inclusion of backflow is
shown in Fig.\ \ref{fig:backflow}.  The energy reduction is greater in the
$\beta$-tin phase than in the diamond phase at the same volume, which we
conjecture is due to the metallic nature of the $\beta$-tin phase.  The energy
reduction increases with the density of the system, as has been found in
studies of the electron gas.\cite{kwon_1998} The use of backflow reduces the
energies of the diamond structure by 0.07--0.1 eV per atom and the $\beta$-tin
structure by 0.11--0.13 eV per atom, over the respective volume ranges, but
the energy difference between the structures at the transition is reduced by
only about 0.015 eV\@.

\begin{figure}
\includegraphics[clip,scale=0.3]{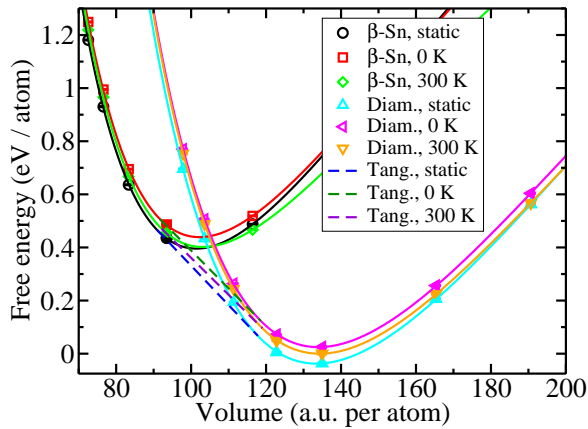}\\
\caption{(Color online) Variation of the free energy with volume for the
  diamond and $\beta$-tin phases of Si calculated within DMC and with
  corrections for the vibrational energy from PBE-DFT\@.  Static lattice,
  $T=0$ and $T=300$ K results are shown, and the zero of free energy is taken
  to be the DMC results including vibrational effects at $300$ K.  The common
  tangents labeled ``Tang.''  join the coexistence points.  Filled symbols
  denote that the structures were found to be dynamically stable within the
  DFT phonon calculations.  The statistical error bars on the DMC energies are
  in the range 0.00005--0.00007 a.u.\ per atom in all cases.
  \label{fig:free-energies_versus_volume}}
\end{figure}

The total Helmholtz free energy at temperature $T$ at each volume was
evaluated as the sum of the DMC energy, the finite-size correction from the
scheme of Kwee \textit{et al.},\cite{Kwee_2008} and the phonon free energy.
The parameters of the Vinet equation of state\cite{vinet_86} were then
calculated using a least-squares fitting procedure.  The Helmholtz free
energies for the static lattice, zero temperature, and $T= 300$ K cases are
shown in Fig.\ \ref{fig:free-energies_versus_volume}.  The fitted
equation-of-state parameters for the diamond and $\beta$-tin structures are
given in Tables \ref{table:diamond_structure_parameters} and
\ref{table:beta-tin_structure_parameters}, respectively, and coexistence
pressures, etc., for the phase transition are reported in Table
\ref{table:transition_pressures}.

\begin{table*}[ht!]
\begin{tabular}{lr@{.}lr@{.}lr@{.}l} \hline \hline

Source & \multicolumn{2}{c}{$V_0$ (a.u.\ / atom)} & \multicolumn{2}{c}{$B_0$
(GPa)} & \multicolumn{2}{c}{$B_0^{\prime}$} \\

\hline

Experiment    & ~~~~~~~$135$&$1$\footnote{Ref.\ \onlinecite{Windisch_1990}.} &
~~$97$&$9(1)$\footnote{Ref.\ \onlinecite{McSkimin_1964}.} & ~$4$&$24^b$ \\

DMC           & $134$&$3(1)$     & $97$&$1(3)$        & $4$&$18(5)$ \\

DMC + BFD-pp  & $132$&$3(1)$     & $99$&$7(3)$        & $4$&$18(5)$ \\

DMC + EMP-pp  & $133$&$3(1)$     & $96$&$2(3)$        & $4$&$19(5)$ \\

DMC + PBE-pp  & $153$&$7(1)$     & $85$&$3(6)$        & $3$&$67(5)$ \\

DMC + LDA-pp  & $131$&$4(1)$     & $101$&$5(3)$       & $4$&$18(5)$ \\

PBE-pp        & $138$&$0$        & $86$&$0$           & $4$&$52$ \\

LDA-pp        & $132$&$8$        & $94$&$4$           & $4$&$46$ \\

\hline \hline
\end{tabular}
\caption{\label{table:diamond_structure_parameters} The equilibrium volume per
  atom $V_0$, bulk modulus $B_0$, and pressure derivative of the bulk modulus
  $B_0^{\prime}$ of diamond-structure Si from experiment  and from DMC and DFT
  calculations.  All results are for 300 K.}
\end{table*}

\begin{table*}[ht!]
\begin{tabular}{lr@{.}lr@{.}lr@{.}l} \hline \hline

Source       & \multicolumn{2}{c}{$V_0$ (a.u.\ / atom)} &
\multicolumn{2}{c}{$B_0$ (GPa)} & \multicolumn{2}{c}{$B_0^{\prime}$} \\

\hline

DMC          & ~~~~~~~$103$&$1(1)$ & $102$&$7(26)$ & $4$&$7(2)$ \\

DMC + BFD-pp & $101$&$4(1)$ & $106$&$2(26)$ & $4$&$7(2)$ \\

DMC + EMP-pp & $102$&$4(1)$ & $104$&$2(26)$ & $4$&$7(2)$ \\

DMC + PBE-pp & $101$&$5(1)$ & $105$&$3(21)$ & $4$&$7(2)$ \\

DMC + LDA-pp & $100$&$6(1)$ & $107$&$4(21)$ & $4$&$8(2)$ \\

PBE-pp       & $104$&$8$    & $94$&$1$      & $4$&$74$   \\

LDA-pp       & $103$&$9$    & $96$&$0$      & $4$&$76$   \\

\hline \hline
\end{tabular}
\caption{\label{table:beta-tin_structure_parameters} The equilibrium volume
  per atom $V_0$, bulk modulus $B_0$, and pressure derivative of the bulk
  modulus $B_0^{\prime}$ of $\beta$-tin-structure Si from DMC and  DFT
  calculations.  All results are for 300 K.}
\end{table*}

\begin{table*}[ht!]
\begin{tabular}{lr@{}lr@{.}lr@{.}lr@{.}lr@{.}l} \hline \hline

Source & \multicolumn{2}{c}{$p_t$ (GPa)} & \multicolumn{2}{c}{$V_{\rm dia}$
(a.u.\ / atom)} & \multicolumn{2}{c}{$V_{\beta-{\rm tin}}$ (a.u.\ / atom)} &
\multicolumn{2}{c}{$\Delta V_t$ (a.u.\ / atom)} & \multicolumn{2}{c}{$\Delta
F_0$ (eV / atom)} \\

\hline

Experiment (Ref.\ \onlinecite{McMahon_1994}) & ~$11$&$.7$        &
~~~~~~~$122$&$48$         & ~~~~~~~~~$94$&$17$        & ~~~~~~~$-28$&$31$
& \multicolumn{2}{c}{-}       \\

LDA (This work)                              & $6$&$.17$        & $125$&$42$
& $95$&$63$        & $-29$&$79$     & ~~~~~~$0$&$2681$    \\

PBE-GGA (This work)                          & $8$&$.75$        & $126$&$82$
& $96$&$95$        & $-29$&$87$     & $0$&$2538$ \\

DMC (This work)                              & $15$&$.1(1)$     & $119$&$0(1)$
& $92$&$1(1)$      & $-26$&$9(1)$   & $0$&$405(3)$ \\

DMC + BFD-pp (This work)                     & $15$&$.7(1)$     & $117$&$1(1)$
& $90$&$5(1)$      & $-26$&$6(1)$   & $0$&$418(3)$ \\

DMC + EMP-pp (This work)                     & $12$&$.2(1)$     & $120$&$3(1)$
& $93$&$2(1)$      & $-27$&$1(1)$   & $0$&$329(3)$ \\

DMC + PBE-pp (This work)                     & $14$&$.9(1)$     & $134$&$1(1)$
& $91$&$0(1)$      & $-43$&$1(1)$   & $0$&$656(3)$ \\

DMC + LDA-pp (This work)                     & $16$&$.9(1)$     & $115$&$6(1)$
& $89$&$3(1)$      & $-26$&$3(1)$   & $0$&$446(3)$ \\

DMC (Ref.\ \onlinecite{alfe_2004_betatin})   & $15$&$.5(5)$     &
\multicolumn{2}{c}{-} &  \multicolumn{2}{c}{-} & \multicolumn{2}{c}{-} &
\multicolumn{2}{c}{-} \\

DMC (Ref.\ \onlinecite{hennig_2010})         & $14$&$(1)$       & $122$&$4(3)$
& $93$&$8(7)$      & $-28$&$69(8)$  & $0$&$42(2)^\dagger$ \\

AFQMC (Ref.\ \onlinecite{Purwanto_2009})     & $12$&$.6(3)$     & $122$&$48$*
& $94$&$17$*       & $-28$&$31$*    & \multicolumn{2}{c}{-}        \\

\hline \hline
\end{tabular}
\caption{\label{table:transition_pressures} The experimental
diamond$\rightarrow \! \beta$-tin phase transition pressure  and volumes and
calculated coexistence pressures and volumes in Si, and the difference between
the zero-pressure free energies.  Volumes are given in a.u.\ per atom. The
asterisks denote that the experimental value was assumed.  The data are for $T
= 300$ K except for the value marked with a dagger, which is for zero
temperature.  The addition of a 300 K temperature correction to the value of
$\Delta F_0$ from Ref.\ \onlinecite{hennig_2010} would result in a decrease of
about 0.024 eV per atom.}
\end{table*}

\section{Testing different pseudopotentials \label{subsec:testing
pseudopotentials}}

Pseudopotentials are very well established in DFT calculations, where they can
give results in excellent agreement with all-electron calculations.  The
accuracy of pseudopotentials in reproducing all-electron results in explicitly
correlated calculations such as QMC is rather more doubtful.  One difficulty
is that there is no straightforward equivalent of the method normally used
within independent electron theories of constructing a pseudopotential by
inverting the atomic Schr\"odinger equation for the valence electrons.
Indeed, the idea of dividing the electrons into core and valence shells is not
well defined in the many-body system.  The difficulties in generating
pseudopotentials which are consistent with correlated methods for calculating
the valence electronic structure leads to additional errors.

To examine the DMC results using different pseudopotentials we calculated
corrections to the energy-volume curves using different pseudopotentials.  We
modeled the interactions between the valence electrons by the PBE-GGA
functional and approximated the DMC energy for pseudopotential X-pp by
\begin{eqnarray}
\label{eq:energy_correction} \tilde{E}^{\rm X-pp}_{\rm DMC} = E^{\rm
DF-pp}_{\rm DMC}  +  \left( E^{\rm X-pp}_{\rm PBE} - E^{\rm DF-pp}_{\rm PBE}
\right) \;.
\end{eqnarray} where the subscript refers to the method used to
calculate the energy, and DF-pp denotes the Dirac-Fock
pseudopotential.\cite{Trail_2005} We tested four pseudopotentials: ultrasoft
LDA and PBE pseudopotentials, the empirical (EMP) pseudopotential of Ref.\
\onlinecite{lee_2000}, and the Hartree-Fock pseudopotential of Burkatzki,
Filippi, and Dolg (BFD) from Ref.\ \onlinecite{Burkatzki_2007}.

There is some evidence that Hartree-Fock (and presumably Dirac-Fock)
pseudopotentials give better results within DMC than DFT
pseudopotentials.\cite{Greeff_1998} The EMP pseudopotential uses the measured
values of the ionization energy of Si$^{+3}$ ions which have a single electron
in the $3s$, $3p$, or $3d$ orbital outside of the neon core, and calculated
data for the amount of valence charge density outside of the pseudopotential
core radius in the ionic states.  Corrections for the core relaxation effects
arising from generating the pseudopotentials in ionized configurations are
also included.  Note that Esler \textit{et al.}\cite{bn_qmc_2010} have
developed a method for calculating a correction to the QMC energy that
accounts for the error due to the pseudopotential approximation. The method
uses all-electron results for small systems, although these would be very
costly to evaluate for Si.

\subsection{Comparison of equation-of-state
parameters \label{subsec:equation_of_state_parameters}}

Data for the equilibrium volume $V_0$, the bulk modulus $B_0$, and the
pressure derivative of the bulk modulus $B_0^{\prime}$ of diamond-structure Si
at 300 K are given in Table \ref{table:diamond_structure_parameters}.  As
shown in numerous DFT calculations, the equilibrium volume $V_0$ of
diamond-structure Si with the PBE functional is larger than experiment while
the LDA value is smaller than experiment.  The PBE value of $B_0$ is
significantly smaller than experiment while the LDA value is close to
experiment, and the PBE and LDA values of $B_0^{\prime}$ are rather larger
than experiment.\cite{Windisch_1990,McSkimin_1964} The experimental value of
$B_0$ is believed to be very reliable and therefore we can be confident that
PBE gives too small a value.  Reference \onlinecite{McSkimin_1964} does not
report an error bar for the measurement of $B_0^{\prime}$ and so it may not be
safe to draw conclusions about the accuracy of experimental or calculated
values for it.

The DMC calculations for the diamond structure with the Dirac-Fock
pseudopotential provide very accurate results for $V_0$ and $B_0$.  The DMC
calculations with corrections for using other pseudopotentials [see Eq.\
(\ref{eq:energy_correction})] give results of variable quality.  The
equation-of-state parameters obtained using the BFD pseudopotential are very
similar to those obtained using the Dirac-Fock pseudopotential.  This is to be
expected because these pseudopotentials were designed to reproduce
Hartree-Fock or Dirac-Fock results, and the relativistic effects included in
the Dirac-Fock pseudopotential are unimportant for Si.  However, the methods
used to construct the pseudopotentials are rather different, and it is
therefore relevant to compare the results obtained with them.  An earlier
comparison of the Trail-Needs Dirac-Fock pseudopotentials and the BFD
pseudopotentials for the LiH molecule also found that very similar results are
obtained with the two classes of pseudopotential.\cite{Trail_2008} The
equation-of-state parameters obtained with the EMP pseudopotential are close
to those from the Dirac-Fock pseudopotential, but the results obtained with
the PBE and LDA pseudopotentials are significantly different and are in poorer
agreement with experiment.  We note that the DMC calculations with the
Dirac-Fock, BFD, and EMP pseudopotentials, which give the best values of $V_0$
and $B_0$, also provide the values of $B_0^{\prime}$ in the best agreement
with experiment (note, however our remark about the lack of an error bar on
the experimental value of $B_0^{\prime}$).  The PBE and LDA pseudopotentials
give results which differ substantially from experiment.

The equation-of-state parameters for $\beta$-tin-structure Si at 300 K are
given in Table \ref{table:beta-tin_structure_parameters}.  Experimental data
for $V_0$, $B_0$, and $B_0^{\prime}$ are not available for the $\beta$-tin
structure of Si.  We note that the values of $V_0$ and $B_0^{\prime}$ for the
$\beta$-tin structure are similar for the DMC-based and DFT-based
calculations, but the DMC-based calculations give larger values of $B_0$.

\subsection{Phase transition pressures and
volumes \label{subsec:phase_transition_pressures_volumes}}

The best experimental value for the room temperature phase transition pressure
is probably that of McMahon and Nelmes of $p_t=11.7$ GPa,\cite{McMahon_1994}
which is close to the value of $p_t=11.3(2)$ GPa found earlier by Hu
\textit{et al.}\cite{Hu_1986} Reference \onlinecite{McMahon_1994} also
reported the structural parameters of the diamond and $\beta$-tin structures
at the transition.  Several factors should be borne in mind when comparing the
experimental and theoretical results.  The experimental transition pressure
and volumes were measured on compression while theoretical studies normally
report the coexistence pressure, which is expected to be lower because
strongly first-order transitions are likely to involve passing over kinetic
barriers.  Unfortunately the diamond$\rightarrow \! \beta$-tin phase
transition in Si is irreversible and different structural phases are found on
decompression,\cite{mujica03} so that the coexistence pressure cannot be
bracketed by the values obtained on compression and decompression.  Prior to
compression, the powdered sample is isotropic, but the compression has a
uniaxial component and, after the transition, the $\beta$-tin phase shows a
``preferred orientation'', that is, the microcrystallites have a tendency to
align, indicating that the stress on the sample during the transition is
anisotropic.  Various
calculations\cite{Libotte_2000,Cheng_2003,Gaal-Nagy_2006a,Durandurdu_2008}
have shown that an anisotropic stress lowers the transition pressure (the
pressure is equal to minus one third of the trace of the stress tensor).  As
the effects of the anisotropic stress and the kinetic barrier act in opposite
directions on the measured transition pressure, and it is difficult to
estimate their size, it seems reasonable to neglect them.  Finally, the
experimental data were measured at room temperature, while the electronic
structure calculations were performed at zero temperature and the vibrational
effects were added afterwards.

The data in Table \ref{table:transition_pressures} show that the DMC
calculations with finite-size and temperature corrections give a coexistence
pressure of 15.1(1) GPa, which is 3.4 GPa larger than the measured transition
pressure and the DMC calculations underestimate the transition volumes.
Correcting the DMC energies to give an approximation to the results that would
be obtained using the EMP potential using Eq.\ (\ref{eq:energy_correction})
gives results for the coexistence pressure and volumes in quite good agreement
with the experimental data.  Correcting the DMC energies to approximate the
results that would be obtained for the PBE and LDA pseudopotentials leads to
poorer transition volumes and pressures.

\section{Conclusions \label{sec:conclusions}}

Backflow improves the description of exchange-correlation effects and
significantly lowers the VMC and DMC energies of the diamond and $\beta$-tin
phases of Si.  These energy reductions, however, mostly cancel in the DMC
energy differences between the two phases and, overall, backflow favors the
$\beta$-tin phase with respect to the diamond phase by roughly 0.015 eV per
atom at the transition.  In a previous DMC study it was suggested that the
overestimation of the experimental transition pressure was most likely to be
due to fixed-node errors.\cite{alfe_2004_betatin} The current much-refined
study includes backflow transformations which reduce the fixed-node error, but
the DMC results with the Dirac-Fock pseudopotential (corrected for finite-size
and vibrational effects) are within error bars of the results of Ref.\
\onlinecite{alfe_2004_betatin}.  The significant reductions in the energies of
the two phases due to the use of backflow make it less likely that the
residual errors arise from the fixed-node approximation.

We tested two different schemes designed to correct for the finite sizes of
the simulation cells.  The finite-size corrections are of order 0.1 eV per
atom, but the maximum difference between the corrections provided by the two
schemes is only about 0.007 eV per atom.  This suggests that both schemes are
working extremely well in this system.  We calculated the phonon frequencies
of the diamond and $\beta$-tin phases as a function of volume, finding that it
is very important to relax the $c/a$ ratio of the $\beta$-tin phase at each
volume to obtain stable phonon modes.  Even then the $\beta$-tin phase was
found to be dynamically stable only over a small volume range.  Our phonon
calculations give a reduction in the transition pressure from including
finite-temperature effects at 300 K of nearly 1 GPa, compared with the
reduction of 1.3 GPa obtained by Ga\'al-Nagy \textit{et al.}\cite{gaal-nagy99}

Finally, we estimated the DMC results that would be obtained with different
pseudopotentials and found a substantial dependence on the pseudopotential.
The Dirac-Fock pseudopotential used in our QMC calculations gave a transition
pressure of 15.1(1) GPa, while the BFD Hartree-Fock pseudopotential gave very
similar results.  The EMP pseudopotential gave 12.2(1) GPa, a PBE
pseudopotential gave 14.9(1) GPa, and an LDA pseudopotential gave 16.9(1) GPa.
The best agreement with experiment was obtained with the EMP pseudopotential.
It is worth noting that the EMP pseudopotential is the only one designed for
use in explicitly correlated calculations, and that it gave the best results
among the Si pseudopotentials tested in correlated calculations in Ref.\
\onlinecite{lee_2000}.  The values of the equilibrium volume, bulk modulus,
and the pressure derivative of the bulk modulus obtained with the EMP
pseudopotential were also in good agreement with experiment.

The results of our QMC calculations for Si and, presumably, calculations with
other atoms, depend significantly on the pseudopotentials used.  We therefore
believe that more research is required into the development of
pseudopotentials for explicitly correlated methods such as QMC\@.

\section*{Acknowledgments} R.M.\ acknowledges financial support
provided by Precursory Research for Embryonic Science and Technology, Japan
Science and Technology Agency (PRESTO-JST), and by a Grant-in-Aid for
Scientific Research in Priority Areas [Development of New Quantum Simulators
and Quantum Design (No.\ 17064016)] (Japanese Ministry of Education, Culture,
Sports, Science, and Technology KAKENHI-MEXT)\@.  N.D.D.\ acknowledges
financial support from the Leverhulme Trust, Jesus College, Cambridge, and the
UK Engineering and Physical Sciences Research Council (EPSRC)\@.  R.J.N.\
acknowledges financial support from EPSRC\@.  We thank John Trail for
assistence with the pseudopotentials.

\end{document}